\newcommand*{\wn}{cm$^{-1}$}
\newcommand*{\Hm}{H$_{2}$}
\newcommand*{\Dm}{D$_{2}$}

\newcommand*{\HDi}{HD$^{+}$}
\newcommand*{\et}{\emph{et al.}}

\def\apj{{\it Astroph.\ J.\ }}

\def\apb{{\it Appl.\  Phys.\  B\ }}

\def\pra{{\it Phys.\  Rev.\ A\ }}
\def\prd{{\it Phys.\  Rev.\ D\ }}
\def\prl{{\it Phys.\  Rev.\ Lett.\ }}

\def\sci{{\it Science\ }}

\def\jms{{\it J. Mol.\ Spectrosc.\ }}
\def\jcp{{\it J.\  Chem.\ Phys.\ }}

\def\cjp{{\it Can.\ J. Phys.\ }}

\def\pccp{{\it Phys.\ Chem.\ Chem.\ Phys.\ }}
\def\jctc{{\it J.\ Chem.\ Theory\ Comp.\ }}
\documentclass[12pt]{iopart}
\usepackage{graphics}
\usepackage{xfrac}
\usepackage{verbatim}

\begin{document}

\title{Constraints on extra dimensions from precision molecular spectroscopy}

\author{E. J. Salumbides$^{1,2}$, A. N. Schellekens$^{3,4,5}$, B. Gato-Rivera$^{3,5}$, and W. Ubachs$^{1}$ }
\address{$^1$Department of Physics and Astronomy, and LaserLaB, VU University, De Boelelaan 1081, 1081 HV Amsterdam, The Netherlands}
\address{$^2$Department of Physics, University of San Carlos, Cebu City 6000, Philippines}
\address{$^3$Nikhef, Science Park 105, 1098 XG Amsterdam, The Netherlands}
\address{$^4$Institute for Mathematics, Astrophysics and Particle Physics, Radboud University, 6500 GL Nijmegen, The Netherlands}
\address{$^5$Instituto de Fisica Fundamental, CSIC, 28006 Madrid, Spain}


\begin{abstract}

Accurate investigations of quantum level energies in molecular systems are shown to provide a test ground to constrain the size of compactified extra dimensions.
This is made possible by the recent progress in precision metrology with ultrastable lasers on energy levels in neutral molecular hydrogen (H$_2$, HD and D$_2$) and the molecular hydrogen ions (H$_2^+$, HD$^+$ and D$_2^+$).
Comparisons between experiment and quantum electrodynamics calculations for these molecular systems can be interpreted in terms of probing large extra dimensions, under which conditions gravity will become much stronger. Molecules are a probe of space-time geometry at typical distances where chemical bonds are effective, i.e. at length scales of an \AA. Constraints on compactification radii for extra dimensions are derived within the Arkani-Hamed-Dimopoulos-Dvali framework, while constraints for curvature or brane separation are derived within the Randall-Sundrum framework. Based on the molecular spectroscopy of D$_2$ molecules and HD$^+$ ions, the compactification size for seven extra dimensions (in connection to M-theory defined in 11 dimensions) of equal size is shown to be limited to $R_7 < 0.6  \mu$m. While limits on compactification sizes of extra dimensions based on other branches of physics are compared, the prospect of further tightening constraints from the molecular method is discussed.

\end{abstract}

\submitto{\NJP}
\maketitle

\section{Introduction}

A standard description of the World is usually presented in terms of the observable $3+1$ spatio-temporal dimensions. At the same time string theories have been developed, seeking to produce a consistent description of the Standard Model of physics including the phenomenon of gravity, which appear to be most consistent if large numbers of dimensions are postulated. A 26-dimensional space-time was deemed necessary for bosonic strings~\cite{Lovelace1971} and a ten-dimensional one for type-II~\cite{Gliozzi1976,Green1981} and heterotic strings~\cite{Gross1985}. The latter theories are closely related to a mysterious theory called M-theory, which lives in 11 dimensions~\cite{Duff1987}.
In contrast, classical physics requires 3 spatial dimensions, e.g. to accommodate Newton's inverse square law as argued already by Immanuel Kant. Ehrenfest has shown that atoms only exhibit stable orbits in a 3-dimensional space~\cite{Ehrenfest1920}. These contradictions between requirements from classical and quantum physics for a 3-dimensional space and the possibility of a theory involving higher dimensions were already resolved in 1926 by Klein invoking the concept of compactification~\cite{Klein1926}.

In the present study the accurate results from precision measurements on molecules are exploited to constrain existing theories on higher dimensions.
For molecular systems, state-of-the-art quantum level calculations of the molecular ions H$_2^+$, HD$^+$, and D$_2^+$, all fundamental three-particle Coulomb systems, have reached the precision that the uncertainty becomes limited by the precision at which values of the fundamental mass ratios $m_p/m_e$ and $m_n/m_p$ are known~\cite{Korobov2009,Karr2011,Korobov2012}, although the recently improved determination of $m_p/m_e$ \cite{Sturm2014} demonstrates active progress on the experimental side.
While experiments on the ro-vibrational spectrum of the H$_2^+$ isotopomer are still under way~\cite{Karr2014} the small dipole moment of the HD$^+$ isotopomer has enabled the accurate study of electric-dipole-allowed transitions in various bands~\cite{Koelemeij2007,Bressel2012,Koelemeij2012}.

In recent years great progress has also been made on the calculation of level energies in the neutral hydrogenic molecules. Accurate Born-Oppenheimer energies have been calculated for the electronic ground state of H$_2$, HD and D$_2$~\cite{Pachucki2010}, as well as non-adiabatic interactions~\cite{Pachucki2008,Pachucki2009} and relativistic and quantum electrodynamical (QED) corrections~\cite{Pachucki2005,Pachucki2007}. Now a full set of ro-vibrational level energies of all quantum states up to the dissociation limit is available for all three isotopomers~\cite{Piszcziatowski2009,Komasa2011}. These calculations on the ground electronic quantum levels were tested in experiments measuring the dissociation limits of H$_2$~\cite{Liu2009}, D$_2$~\cite{Liu2010}, and HD~\cite{Sprecher2010}. Further they were compared to experimental values for the fundamental vibrational splitting in H$_2$ and the hydrogen isotopomers~\cite{Dickenson2013,Niu2014}, to a measurement of the first overtone in H$_2$~\cite{Campargue2012,Kassi2014} and D$_2$~\cite{Kassi2012}, a measurement of the second overtone in H$_2$~\cite{Hu2012,Tan2014}, and measurements of highly excited rotational levels in H$_2$~\cite{Salumbides2011}. The results from the variety of experimental precision measurements on both the ionic and neutral hydrogen molecules are generally in excellent agreement with the QED-calculations, within combined uncertainty limits from theory and experiment.

The agreement between experiment and first-principles calculations on quantum level energies of molecules has inspired an interpretation of these data that goes beyond molecular physics. Since weak, strong and (Newtonian) gravitational forces have negligible contributions to their quantum level structure, electromagnetism is the sole force acting between the charged particles within light molecules, and QED is the fully-encompassing framework to perform calculations. This makes it possible to derive bounds on possible fifth forces between hadrons from molecular precision experiments compared with QED-calculations~\cite{Salumbides2013,Salumbides2014}.

Theories of higher dimensions were developed with the goal to resolve the hierarchy problem, i.e. the vast difference of scales between that of electro-weak unification (1 TeV) and that of the Planck scale (10$^{16}$ TeV), where gravity becomes strong. By permitting the leakage of gravity into higher dimensions while keeping the particles and the three forces of the Standard Model in 3+1 dimensions, and invoking a compactification range for the extra dimensions exceeding 3+1, two different testable theories were phrased by Arkhani-Hamed, Dimopoulos, and Dvali~\cite{Hamed1998} and by Randall and Sundrum~\cite{Randall1999,Randall1999b}. The mathematical formalisms of these theories can be applied to molecular physics test bodies, from which constraints on the compactification distances can be deduced for the former,
while constraints on the brane separation or curvature can be derived for the latter. That is the subject of the present paper.

\section{The ADD-model}
\label{ADD-model}
It is the intention of the theory formulated by Arkani-Hamed, Dimopoulos, and Dvali~\cite{Hamed1998}, referred to as ADD-theory, to establish an effective Planck scale to coincide with the electro-weak scale by allowing gravity to propagate in extra dimensions. The three forces of the Standard Model, tested at very short distances in particle and atomic physics experiments, are considered to act locally within a 3-brane (3 spatial dimensions and a time dimension) embedded in a higher dimensional bulk, where gravity may act allowing for gravitons to escape. By this means in ADD the hierarchy problem is nullified, and the so-called desert range between the electro-weak scale ($M_\mathrm{EW}$) of 1 TeV and the Planck scale ($M_\mathrm{Pl}$) of 10$^{16}$ TeV avoided. The extension of the extra dimensions is necessarily limited, in the case of flat metrics considered in ADD, since experiments of the Cavendish-type have proven that gravity obeys the Newtonian $1/r$ potential beyond the range of 1 cm \cite{Adelberger2003}. Hence, the extra dimensions are considered to be compactified within a range parameter $R_n$. While in principle the extra dimensions could exhibit differing range parameters, in the ADD-formalism and in the present analysis such difference is not made.

The Newtonian gravitational potential may be written as:
\begin{equation}
  V_\mathrm{N}(r)=  -G\frac{m_1m_2}{r} = -\frac{m_1m_2}{M_\mathrm{Pl}^2} \frac{1}{r} \hbar c
\end{equation}
with the Planck mass defined as $M_{Pl}^2=\hbar c/G$ in SI units.
In the following discussions, we adopt the natural units $\hbar=c=1$ and drop the $(\hbar c)$-factor in the potentials.
The extra $n$ spatial dimensions proposed in the ADD theory result in a modification of Newtonian gravity, for distances shorter than the compactification length range, that is consistent with Gauss law:
\begin{equation}
  V_\mathrm{ADD} =  -\frac{m_1m_2}{ M^{n+2}_{(4+n)}} \frac{1}{r^{n+1}},
\label{ADD-short}
\end{equation}
where the subscript 4 represents the known $(3+1)$ spacetime dimensions, and $M_{(4+n)}$ is the full higher-dimensional Planck mass.
For separations larger than the compactification length, $r > R_n$, the ADD potential should be in correspondence with the Newtonian $1/r$-form
\begin{equation}
  V_\mathrm{ADD} = -\frac{ m_1 m_2 }{ M^{n+2}_{(4+n)}(R_n )^n } \frac{1}{r}.
\label{ADD-long}
\end{equation}
To be more precise, $(R_n )^n$ should be the compactified volume of the extra dimensions $V_n$, thus a factor of order unity might be included for a specific compactification geometry.

The Planck mass $M_\mathrm{Pl}$ is then related to the higher-dimensional mass $M_{(4+n)}$ via:
\begin{equation}
  M_\mathrm{Pl}^2 =  M^{n+2}_{(4+n)} (R_n)^n.
\label{Fund_M_Pl}
\end{equation}
Thus the fundamental mass $M_{(4+n)}$ may still be small and $M_\mathrm{Pl}$ becomes large due to the compactified volume of extra dimensions.
Arkani-Hamed~\et\ have shown that if the fundamental mass is taken as $M_\mathrm{EW}$ one extra dimension would have a range of order 10$^{10}$ km to account for the weakness of gravity. This is incompatible with experimental evidence. But for two extra dimensions $R_n$ would be of sub-millimeter size~\cite{Hamed1998}, thus at a range where Newtonian gravity is not firmly tested.
In our present study we will not set a certain energy scale, and in particular we do not assume that $M_{4+n}\sim M_\mathrm{EW}$. Our goal is to constrain $R_n$ from molecular physics experiments without theoretical prejudice regarding the fundamental mass scale.

While dealing with molecules the unit attraction of gravity can be chosen as that between two protons and a dimensionless gravitational coupling strength is defined as:
\begin{equation}
 \alpha_G=Gm_p^2/\hbar c
\end{equation}
Note that this particular choice of the gravitational coupling constant is equivalent to specifying $\alpha_G = (m_p/M_\mathrm{Pl})^2 = 5.9\times10^{-39}$.
Then the Newtonian attraction between two particles consisting of $N_1$ and $N_2$ protons or neutrons ($m_n\simeq m_p$ is adopted) can be written as:
\begin{equation}
  V_\mathrm{N}(r)= -\alpha_G N_1N_2 \frac{1}{r}
\end{equation}
From Eq.~(\ref{Fund_M_Pl}), the ADD-potential of Eq.~(\ref{ADD-short})  within the compactification radius $r < R_n$ may be rewritten as:
\begin{equation}
  V_\mathrm{ADD}(r)=  -\alpha_G N_1N_2 R_n^n \frac{1}{r^{n+1}},
\label{V_ADD}
\end{equation}
while this potential reduces to normal Newtonian gravity $V_\mathrm{N}$ for the range outside the compactification length range $r > R_n$.

\begin{figure}
\begin{indented}
\item[]\resizebox{0.8\textwidth}{!}{\includegraphics{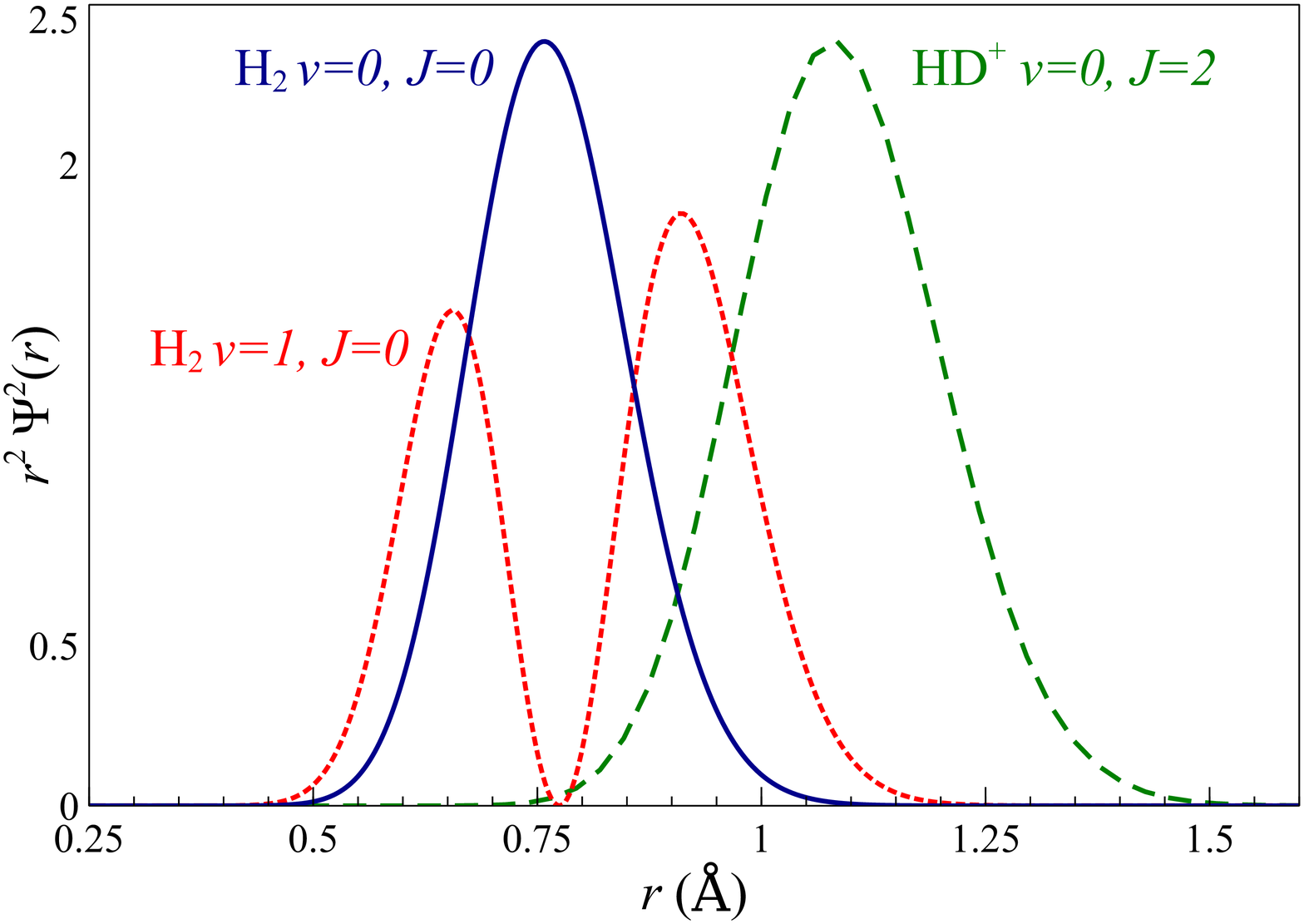}}
\caption{\label{Wavefunctions} Wave functions for H$_2$ in the electronic ground state with $v=0, J=0$ and $v=1,J=0$, and for the HD$^+$ $v=0, J=2$ quantum state.}
\end{indented}
\end{figure}

For molecules this gravitational potential has an effect on the level energy of a molecular quantum state with wave function $\Psi(r)$, to be written as an expectation value:
\begin{eqnarray}
 \left<V_\mathrm{ADD}\right> &=& -\alpha_G N_1N_2 \left[ \quad \int_{R_n}^{\infty} \Psi^*(r) \frac{1}{r} \Psi(r) r^2 dr \right.  \nonumber \\
 &&\left. \quad\quad\quad\quad\quad +R_n^n \int_0^{R_n} \Psi^*(r) \frac{1}{r^{n+1}} \Psi(r) r^2 dr \quad \right]
\label{Integrals}
\end{eqnarray}
Note that the wave functions are given along a single coordinate $r$, i.e. the vibrational coordinate, that probes the gravitational forces between nucleons.
Here the nuclear displacement is separated from electronic motion and the wave function $\Psi(r)$ represents the probability that the nuclei in the molecule are at internuclear separation $r$.
The first integral term represents the ordinary gravitational attraction, which is for protons $8 \times 10^{-37}$ times weaker than the electrostatic repulsion, and can therefore be neglected.
The second integral represents the effect of modified gravity and is evaluated using accurate wave functions for H$_2$.
The wave functions of the H$_2$ ground electronic state for the $v=0$ and 1 levels are shown in Fig.~\ref{Wavefunctions}. In practice, the integration is performed up to $r=10$ \AA\ since the wave function amplitude is negligible beyond that. Also at shorter distances $r < 0.1$ \AA\ the wave function amplitude becomes negligible, for which reason the second integral in Eq.~(\ref{Integrals}) converges without additional assumptions.
The HD$^+$ $v=0, J=2$ ground electronic state wave function is also displayed in Fig.~\ref{Wavefunctions} showing the larger internuclear distance of the ion with respect to the neutral.

For transitions between quantum states $\Psi_1$ and $\Psi_2$, as in spectroscopic transitions in molecules, a differential effect must be calculated:
\begin{equation}
  \left<\Delta V_\mathrm{ADD}(n,R_n) \right> =  -\alpha_G N_1N_2 R_n^n
 \left[ \left< {\Psi_1} \left| \frac{1}{r^{n+1}} \right|{\Psi_1} \right> - \left< {\Psi_2} \left| \frac{1}{r^{n+1}} \right|{\Psi_2} \right> \right]
\label{ADD-trans}
\end{equation}
This equation represents the expectation value for a high-dimensional gravity contribution to transitions in molecules. Here the ADD-expectation value is written explicitly as a function of the two relevant parameters:
the number $n$ of extra spatial dimensions and the compactification scale $R_n$.
From Eq.~(\ref{ADD-trans}), it is clear that a stronger effect can be expected if the difference in wave functions of the two states $\Psi_1$ and $\Psi_2$ is greater. For this reason, measurements on the dissociation limit in molecules, where $\Psi_1$ is lowest energy bound state and $\Psi_2$ is the non-interacting two-atom limit at $r=\infty$, are the most sensitive probes.

\section{The Randall-Sundrum models}

Let us now consider the Randall-Sundrum scenarios, RS-I and RS-II, to approach the physical description of extra dimensions in an alternative manner~\cite{Randall1999,Randall1999b}. 
In these scenarios, the particles and interactions of the Standard Model are confined in the SM-brane, separated by some distance $y_c$ from another (hidden) 3-brane along \emph{one} extra dimension $y$.  
The branes and the bulk are sources of gravity that were shown to produce an Anti-de-Sitter metric:
\begin{equation}
ds^2 = e^{-2 k |y|} \eta_{\mu\nu}dx^\mu dx^\nu  + dy^2,
\label{RS_metric}
\end{equation}
where $\exp(-k|y|)$ is a so-called warp factor and $k$ is the bulk curvature~\cite{Randall1999}.
The warped metric differentiates the RS models from the ADD model with a flat metric where $k=0$.
Thus, the exponential warp factor in the RS scenarios solves the hierarchy problem alternatively, without requiring large extra dimensions as assumed in the ADD model.

In the RS scenarios the modified gravitational potential between two masses separated by a distance $r$ in the SM-brane can be expressed as:
\begin{equation}
	V_\mathrm{RS}(r) =  -G \frac{m_1m_2}{r} \left( 1 + \Delta_\mathrm{RS} \right),
\end{equation}
where $\Delta_\mathrm{RS}$ is the correction to the Newtonian potential.
Callin~\cite{Callin2004b} computed the potential in the framework of the RS-I scenario, obtaining for short distances:
\begin{equation}
  \Delta_\mathrm{RS-I}(r) \simeq
     \frac{4}{3\pi kr}\frac{1-e^{-2k y_c}}{1+\frac{1}{3}e^{-2k y_c}},  \quad kr \ll 1 .
\label{RSI_potential_short}
\end{equation}
Here one can distinguish two regimes, $ky_c \ll 1$  and  $ky_c \gg 1$, with the result up to the leading order:
\begin{equation}
\renewcommand{\arraystretch}{2}
  \Delta_\mathrm{RS-I}(r) \simeq \left\{
  \begin{array}{l l}
   \frac{2y_c}{\pi r} + ..., & \quad ky_c \ll 1,\\
   \frac{4}{3 \pi kr} + ..., & \quad ky_c \gg 1.
  \end{array} \right.
\renewcommand{\arraystretch}{1}
\label{RSI_potential_ky}
\end{equation}
It turns out that the RS potential for long distances ($kr \gg 1$) is not applicable to molecules and is not considered further.

In the RS-II scenario, the hidden 3-brane is chosen to be infinitely far ($y_c \rightarrow \infty$) from the SM-brane resulting in an effective model with a single 3-brane (SM-brane) in the bulk.
This solution thus offers the existence of extra dimensions that do not require compactification in contrast to the ADD model.
For short distances in the RS-II scenario, Callin and Ravndall \cite{Callin2004a} obtained
\begin{equation}
  \Delta_\mathrm{RS-II}(r) = \frac{4}{3\pi kr} + ..., \quad kr \ll 1 ,
  \label{RS_potential_leading_order}
\end{equation}
for the RS correction.
Note the correspondence of Eq.~(\ref{RS_potential_leading_order}) with that of Eq.~(\ref{RSI_potential_ky}) for $ky_c \gg 1$, which is expected since the latter RS-I condition implies the transition to RS-II at infinite brane separation.

From these RS potential corrections, the expectation values of the leading-order shifts of transitions in molecules, in the short distance separation ($kr \ll 1$)  regime, are therefore:
\begin{equation}
 \left<\Delta V_\mathrm{RS}(k) \right> =  \alpha_G N_1N_2 \mathcal{F} \left(\frac{4}{3\pi k}\right) \left[ \left< {\Psi_1} \left| \frac{1}{r^2} \right|{\Psi_1} \right> - \left< {\Psi_2} \left| \frac{1}{r^2} \right|{\Psi_2} \right> \right],
\label{RS-trans-short}
\end{equation}
where $\mathcal{F}=(1-e^{-2k y_c})/(1+\frac{1}{3}e^{-2k y_c})$ for RS-I and $\mathcal{F}=1$ for RS-II.
Using these expressions, limits on the curvature $k$ or the brane separation $y_c$ based on molecular spectroscopy data can be derived.

\section{Constraints on higher dimensions from molecular data}

In the previous section, the expectation value for a higher-dimensional gravity contribution to a transition frequency in a molecule was presented for both ADD and RS approaches to higher dimensions.
This expectation value is interpreted as a contribution to the binding energy of molecules in certain quantum states.
This rationale will be used to derive constraints on characteristic parameters underlying the extra-dimensional theories, the compactifictaion range $R_n$ for the ADD scenario and the warp factor $k$ or brane separation $y_c$ for the RS scenario(s).

In Table~\ref{data} a compilation is made of a comparison between theoretical and experimental values obtained in recent experiments for hydrogen neutral molecules and hydrogen molecular ions, and the stable isotopomers containing deuterons.
Ro-vibrational transitions in the ground electronic state are indicated by the change in vibrational quantum number $v$, while $D_0$ denotes the dissociation energy of the ground electronic state.
In the Table the agreement between theory and experiment is represented by the combined uncertainty $\delta E$ with:
\begin{equation}
 \delta E = \sqrt{\delta E_\mathrm{exp}^2 + \delta E_\mathrm{theory}^2},
\label{Uncertainties}
\end{equation}
where $\delta E_\mathrm{exp}$ and $\delta E_\mathrm{theory}$ signify uncertainties of theory and experiment.
On all but two cases the values for $\delta E$ were found to be larger than the discrepancies between theory and experiment, denoted by $\Delta E = E_\mathrm{exp} - E_\mathrm{theory}$, while the \Hm\ $v=0\rightarrow1$ is within two standard deviations ($\Delta E < 2\,\delta E$). From these results it is concluded that QED-theory for these molecular systems is in very good agreement with observations.
Recent calculations by Korobov \emph{et al.}~\cite{Korobov2014} result in an increased discrepancy with the experimental results of Bressel \emph{et al.}~\cite{Bressel2012} at the level of 2.6 standard deviations, and we do not include the \HDi\ $v=0\rightarrow1$ values in the comparisons.

\begin{table}
\caption{\label{data} Data from recent precision measurements of vibrational energy splittings as well as the dissociation energy $D_0$ in neutral and ionic molecular hydrogen and their isotopomers. Adapted from Ref.~\cite{Salumbides2013} and updated with most recent data. $\Delta E$ represents the deviation between theory and experiment, while  $\delta E$ represents the combined uncertainties, cf. Eq.~(\ref{Uncertainties}).
}
\begin{indented}
\lineup
\item[]\begin{tabular}{@{}l@{\hspace{10pt}}c@{\hspace{15pt}}r@{.}l@{\hspace{15pt}}r@{.}l@{\hspace{15pt}}c}
\br
species & transition & \multicolumn{2}{l}{$\Delta E$ (\wn)} & \multicolumn{2}{l}{$\delta E$ (\wn)} & Ref. \\
\mr
\Hm	&$v=0\rightarrow1$ 	&0&000\,24	&0&000\,17	&\cite{Dickenson2013,Niu2014} \\
	&$v=0\rightarrow2$	&0&000\,4 	&0&002\,0	&\cite{Campargue2012,Kassi2014} \\
	&$v=0\rightarrow3$	&-0&000\,6	&0&002\,5	&\cite{Hu2012,Tan2014} \\
	&$D_0$			&0&000\,0	&0&001\,2	&\cite{Liu2009} \\
\mr
HD	&$v=0\rightarrow1$ 	&0&000\,11	&0&000\,23	&\cite{Dickenson2013,Niu2014} \\
	&$D_0$			&0&000\,9	&0&001\,2	&\cite{Sprecher2010} \\
\mr
\Dm	&$v=0\rightarrow1$ 	&-0&000\,02	&0&000\,17	&\cite{Dickenson2013,Niu2014} \\
	&$v=0\rightarrow2$	&-0&000\,5	&0&001		&\cite{Kassi2012} \\
	&$D_0$			&0&000\,5	&0&001\,1	&\cite{Liu2010} \\
\mr
\HDi	&$v=0\rightarrow1$ 	&-0&000\,005\,2	&0&000\,002\,0 	&\cite{Bressel2012, Korobov2014} \\
	&$v=0\rightarrow4$ 	& 0&000\,009	&0&000\,017	&\cite{Koelemeij2007} \\
\br
\end{tabular}
\end{indented}
\end{table}

The agreement between theory and experiment for molecular systems is now translated into a constraining relation for higher dimensions in the ADD framework:
\begin{equation}
 \left<\Delta V_\mathrm{ADD}(n,R_n)\right> \, < \delta E .
\label{ADD-constrain-relation}
\end{equation}
As a first example we take the measurement on the fundamental vibration in the H$_2$ molecule. This is one of the most accurately measured numbers in neutral molecules, while also the QED-calculations for this fundamental rotationless transition are more accurate by an order of magnitude with respect to the absolute binding energies, because of cancellation of errors for non-rotating molecules~\cite{Dickenson2013}.
Constraints on $R_n$ can be derived via:
\begin{equation}
 (R_n)^n < \frac{\delta E}{\alpha_G  N_1N_2 \Delta}
\label{Rn-constraint}
\end{equation}
with $\Delta$ the difference in expectation values over the wave function densities between $v=0$ and $v=1$ vibrational states in the molecule:
\begin{equation}
 \Delta = \left[ \left< \frac{1}{r^{n+1}} \right>_{\Psi_1} - \left< \frac{1}{r^{n+1}} \right>_{\Psi_0} \right]
\end{equation}

\begin{figure}
\begin{indented}
\item[]\resizebox{0.8\textwidth}{!}{\includegraphics{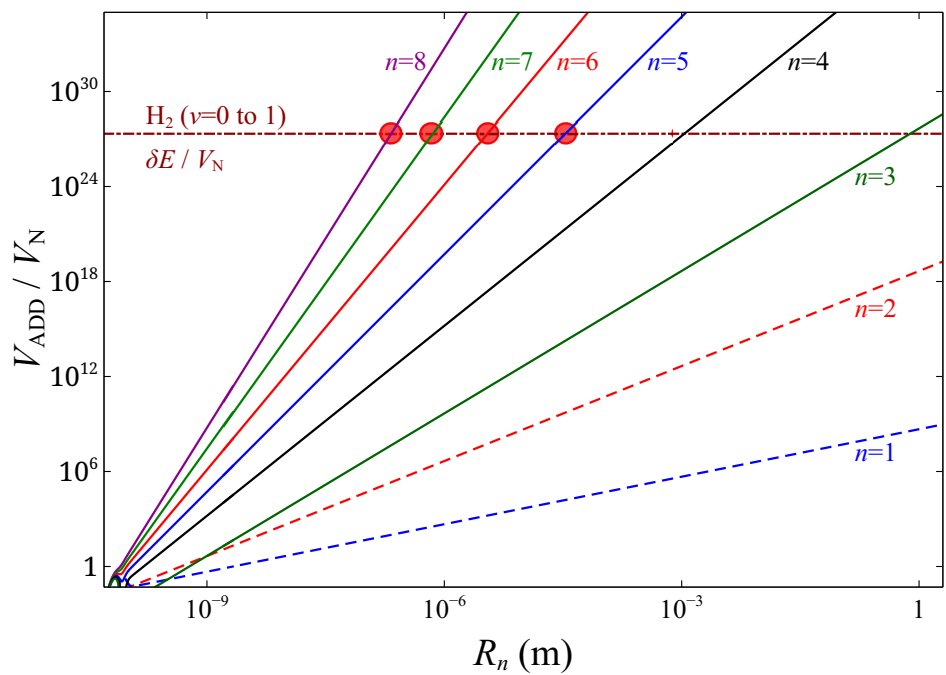}}
\caption{(Color online) Limit on the compactification range $R_n$ as derived from the measurement of the fundamental vibration in the H$_2$ molecule~\cite{Dickenson2013} in comparison with the ADD-formalism.}
\label{Limits-H2}
\end{indented}
\end{figure}

The wave functions for the lowest vibrational states, in the case of H$_2$ and for $J=0$, as obtained from \emph{ab initio} calculations~\cite{Piszcziatowski2009,Komasa2011} are plotted in Fig.~\ref{Wavefunctions}. Since the wave functions are located in the same region of space the fundamental vibrational transition in the hydrogen molecule ($v=0 \rightarrow v=1$) probes only a differential effect.
The resulting constraints on $R_n$ from the measurement of the fundamental vibration in the H$_2$ molecule
for the range of extra dimensions $n=1-8$ are presented in Fig.~\ref{Limits-H2}.
The sloping lines in Fig.~\ref{Limits-H2} represent calculated $V_\mathrm{ADD}/V_\mathrm{N}$ for different $n$ and  $R_n$ values.
The horizontal dashed line $\delta E/V_\mathrm{N}$ indicates limits from molecular spectroscopy.
Hence, for certain numbers of extra dimensions $n$, $R_n$ is constrained to be less than the value where the $V_\mathrm{ADD}/V_\mathrm{N}$ and $\delta E/V_\mathrm{N}$ intersect in the graph.
Constraints on $R_n$, obtained from a comparison with the fundamental vibrational transition of H$_2$ are presented in Table~\ref{Constraints-R}.

\begin{table}
\caption{\label{Constraints-R}Constraints on the size $R_n$ of compactified dimensions (in units of m) as derived from a number of molecular features: (i) the fundamental ($0\rightarrow1$) vibration in H$_2$, (ii) the dissociation limit $D_0$ of H$_2$, (iii) the dissociation limit of D$_2$, and (iv) the (4-0) R(2) ro-vibrational transition in HD$^+$.  The constraints are derived within the ADD-framework assuming that $n$ extra dimensions are of equal size. The corresponding higher-dimensional Planck length $R_{\mathrm{Pl},(4+n)}$ (in units of m) and Planck mass $M_{(4+n)}$ (in units of GeV) are also tabulated, where the smallest values for $R_{\mathrm{Pl},(4+n)}$ and the highest value for $M_{(4+n)}$ is taken from the examples.}
\footnotesize
\begin{indented}
\lineup
\item[]\begin{tabular}{@{}l@{\hspace{17pt}}l@{\hspace{12pt}}l@{\hspace{12pt}}l@{\hspace{12pt}}l@{\hspace{20pt}}l@{\hspace{12pt}}l}
\br
 $n$ & \multicolumn{4}{c}{$R_n$} &  $R_{\mathrm{Pl},(4+n)}$ & $M_{(4+n)}$ \\
\mr
  & H$_2$ (1-0) & H$_2$ $D_0$ & D$_2$ $D_0$ & HD$^+$ (4-0) & (m) & (GeV)  \\
\mr
2 & $2.2\times10^{ 4}$ & $1.0\times10^{ 4}$ & $4.8\times10^{ 3}$ & $2.8\times10^{ 3}$ & $2.1\times10^{-16}$  & $9.3\times10^{-1}$\\
3 & $7.7\times10^{-1}$ & $1.9\times10^{-1}$ & $1.2\times10^{-1}$ & $1.0\times10^{-1}$ & $3.0\times10^{-15}$  & $6.5\times10^{-2}$\\
4 & $1.1\times10^{-3}$ & $8.5\times10^{-4}$ & $5.9\times10^{-4}$ & $7.0\times10^{-4}$ & $1.8\times10^{-14}$  & $1.1\times10^{-2}$\\
5 & $3.3\times10^{-5}$ & $3.2\times10^{-5}$ & $2.4\times10^{-5}$ & $3.1\times10^{-5}$ & $5.8\times10^{-14}$  & $3.4\times10^{-3}$\\
6 & $3.4\times10^{-6}$ & $3.7\times10^{-6}$ & $2.9\times10^{-6}$ & $3.0\times10^{-6}$ & $1.4\times10^{-13}$  & $1.4\times10^{-3}$\\
7 & $6.9\times10^{-7}$ & $7.8\times10^{-7}$ & $6.4\times10^{-7}$ & $6.3\times10^{-7}$ & $2.8\times10^{-13}$  & $7.1\times10^{-4}$\\
\br
\end{tabular}
\end{indented}
\end{table}

The experimental as well as the theoretical results for the fundamental vibration in the hydrogen molecule are known to the 10$^{-4}$ \wn\ level, an order of magnitude more accurate than the values for the binding energies~\cite{Dickenson2013,Niu2014}. However, for a comparison of dissociation limits it is no longer a small difference along the internuclear coordinate axis that is probed, but the difference between the 1 \AA\ molecular scale and infinite atomic separation. The expectation value for the ADD-contribution to the binding energy of the lowest bound state in the H$_2$ molecule, or the $D_0$ binding energy, is:
\begin{equation}
 \left<\Delta V_\mathrm{ADD}(n,R_n)\right> =  \alpha_G  N_1N_2 R_n^n \left< \frac{1}{r^{n+1}} \right>_{\Psi_0}
\end{equation}
By comparing to the experimental findings on $D_0$(H$_2$) \cite{Liu2009} this leads to another set of constraints on $R_n$ for $n$ extra dimensions, which are also listed in Table~\ref{Constraints-R}.

The method was further applied to the fundamental vibration of HD and D$_2$, where the experimental and theoretical uncertanties are similar to those in H$_2$. Although the heavier masses of the isotopomers improve the constraints obtained from H$_2$, as expected from Eq.~(\ref{Rn-constraint}), the HD and D$_2$ fundamental vibration constraints are still less stringent compared to that from the H$_2$ dissociation limit. 
The results obtained for D$_2$ dissociation energy~\cite{Liu2010} lead to the tightest constraints on $R_n$ from the neutrals as listed in Table~\ref{Constraints-R}, which scale by a factor $(\sfrac{1}{4})^{1/n}$ relative to H$_2$ due to the mass difference.

The experimental accuracy for the HD$^+$ molecular ion transitions is an order magnitude better than the corresponding neutral molecule system that stems mostly from the possibility of trapping the ionic species.
The theoretical calculation for the three-body HD$^+$ level energies is also more accurate than those of the neutral molecular hydrogen.
However, the internuclear separation of HD$^+$ ($\sim1.1$ \AA) is greater than that of neutral hydrogen molecules ($\sim0.76$ \AA) as shown in Fig.~\ref{Wavefunctions}.
Thus the neutrals are inherently more sensitive as the wave functions probe shorter internuclear distances compared to their ionic counterparts.
The constraints for $R_n$ derived from the HD$^+$ ($v=0,J=2\rightarrow v=4,J=3$) ro-vibrational transition from Koelemeij et al.~\cite{Koelemeij2007} are listed in Table~\ref{Constraints-R}.
In the table, the $R_n$ constraints from D$_2$ $D_0$ are the most stringent for $n=4,5,6$ extra dimensions while the constraints from HD$^+$ are the most constraining for $n=2,3,7$. 
The higher-dimensional Planck mass $M_{(4+n)}$ and corresponding Planck length $R_{\mathrm{Pl},(4+n)}$ derived from the tightest $R_n$ constraints obtained in this study are also listed in Table~\ref{Constraints-R}.

Similarly, we derive constraints pertaining to corrections in the RS scenario with one extra dimension, and the combined uncertainty $\delta E$ for a specific molecular transition
\begin{equation}
 \left<\Delta V_\mathrm{RS}(k) \right> < \delta E.
\end{equation}
Using the combined uncertainties for the $D_0$(D$_2$) study, we present constraints for the RS schemes.
For the RS-I scenario in the short distance ($kr \ll 1$) regime, we obtain constraints for the brane separation of $y_c<1\times 10^{18}$ m in the limit $ky_c\ll 1$ using Eq.~(\ref{RSI_potential_ky}). In the limit $ky_c\gg 1$ in Eq.~(\ref{RSI_potential_ky}), a constraint for the inverse of the curvature of $1/k < 2\times 10^{18}$ m is obtained.
For the RS-II model, we obtain constraints for the inverse of the curvature $1/k < 2\times 10^{18}$ m for $kr \ll 1$ from Eq.~(\ref{RS-trans-short}).

\section{Comparison with other constraints}
The constraints obtained from molecular systems probe length scales in the order of Angstroms.
This complements bounds probing subatomic to astronomical length scales obtained from other studies using distinct methodologies.
Length scales of several hundred nanometers to microns are probed in Casimir-force studies using cantilevers~\cite{Zuurbier2011} or atomic-force microscopy~\cite{Banishev2013}.
The micrometer to millimeter range are probed in torsion-balance type experiments, with the tightest constraint obtained by Kapner \emph{et al.}~\cite{Kapner2007} for a single extra dimension of $R_1 < 4.4 \times 10^{-5}$ m.
The centimeter to meter separations are accessed by Cavendish- or E\"{o}tv\"{o}s-type investigations in the laboratory, while astronomical scales can be probed in satellite or planetary orbits that also serve to constrain the universality of free fall and deviations from the gravitational inverse-square law~\cite{Adelberger2003}.
Constraints for the RS-theories are obtained by Iorio~\cite{Iorio2012} using data from orbital motions of satellites or astronomical objects, with the tightest constraint for the inverse of the curvature of $1/k < 5$ m obtained from the motion of the GRACE satellite.
The latter constraint is in the $kr \gg 1$ regime of Eq.~(\ref{RS_potential_leading_order}) and probes a different distance range to that of molecules ($kr \ll 1$).

Precision spectroscopies of hydrogen~\cite{Biraben2009,Parthey2010} and muonic atoms~\cite{Pohl2010,Antognini2013} have been interpreted along the same lines in terms of the ADD-model~\cite{Li2007} resulting in typical constraints of $R_3 < 10^{-5}$ m.
The interpretation is not straightforward  because of the proton size puzzle~\cite{Pohl2013}; in fact, the argument has been turned around, where the existence of extra dimensions are instead invoked as a possible solution to the puzzle~\cite{Wang2013}. 
In the treatment of atoms, some assumptions had to be made on the wave function density at $r=0$, typical for the $s$-states involved, causing problems in calculating the second integral of Eq.~(\ref{Integrals}) over the \emph{electronic} wave function that has a significant wave function amplitude at $r=0$ in atoms.
Note that these difficulties are absent in molecules, as the molecular wave function probes the 0.1 - 5 Angstrom distance range.

To probe length scales in the subatomic range, one is ultimately limited by the increasing contributions from nuclear structure and the strong interaction, e.g. in neutron scattering studies~\cite{Nesvizhevsky2008}.
In contrast to QED calculations, the most accurate lattice-QCD calculation of light hadron masses only achieves relative accuracies in the order of a few percent.
Nevertheless, the smaller nucleon size presents higher sensitivity to effects of ADD-type interactions, and constraints for the size of extra dimensions may be extracted.
The general method for molecules presented here may be applied to a comparison of \emph{ab initio} lattice-QCD calculations with the measurements of light hadron masses.
The corresponding QCD test probes length scales of the size of a nucleon at $\sim 10^{-15}$ m.
The \emph{ab initio} calculations of D{\"u}rr \emph{et al.}~\cite{Durr2008} for the nucleon mass are estimated to be accurate to around 50 MeV/$c^2$ while the experimental mass values are accurate to 20 eV/$c^2$.
The calculated nucleon mass, with 3\% relative accuracy, is the isospin average of proton $m_p$ and neutron $m_n$ masses, while $m_p$ is known experimentally to be $0.1\%$ smaller than $m_n$.
Constraints based on experimental nucleon masses and QCD theory have not been explored, but we produce here a rough first estimate by assuming that the three constituent quarks each have an effective mass that is $\sfrac{1}{3}$ of the nucleon mass, and have separation distances $\sim r_p$.
Analogous to Eq.~(\ref{ADD-constrain-relation}) for QED interactions in molecules, the expectation value for an ADD contribution on the mass of the proton can be written as $V_\mathrm{ADD}(p)/c^2 < \delta m_p$, yielding a bound for the case of $7$ extra dimensions of $R_7 < 2.4\times10^{-10}$ m.

In high-energy particle collisions, higher-dimensional gravitons may be produced that could escape into the bulk, leading to events with missing energy in (3+1)-dimensional spacetime~\cite{Hamed1998,Mirabelli1999,Giudice1999}.
Based on this premise of an energy loss mechanism the phenomenology of the SN 1987A supernova was investigated, imposing limits on extra dimensions of $R_2 < 3\times 10^{-6}$ m,  $R_3 < 4\times 10^{-7}$ m, and $R_4 < 2\times 10^{-8}$ m~\cite{Cullen1999}.
Similarly, from a missing energy analysis of proton colliding events at LHC, a constraint for $R_2<3.2\times10^{-4}$ m can be extracted from the $M_{4+n} = 1.93$ TeV bound for $n=2$ given in Ref.~\cite{Aad2013}.
For comparison, the Planck energy scale in $(4+n)$ dimensions in Table~\ref{Constraints-R} turns out to be in the range between $1-1000$ MeV, but are derived from a completely independent methodology.  
Also for $n>2$, the bounds derived from LHC are nominally more stringent than those from molecules.
However, additional assumptions beyond the ADD potential in Eq.~(\ref{ADD-short}), e.g. the fundamental quantization of gravity, the existence and propagation of gravitons in ($4+n$) dimensions and postulating the existence of massive new particles, are necessary for an effective theory~\cite{Mirabelli1999,Giudice1999} to interpret the LHC missing energy signals. Such assumptions are not
needed for the molecular physics bounds, which are not sensitive to physics at very short distances.

\section{Conclusion and outlook}

The alternative approaches to constraining compactification radii for extra dimensions, partially surveyed here, are all complementary as they probe different length and energy scales.
Some approaches serve to produce tighter limits, however, often at the expense of additional assumptions.
In the present study, a constraint is derived on compactification scales of extra dimensions from precision measurements on molecules, leading to straightforward interpretations.
Molecules, in particular the lightest ones as neutral and ionic molecular hydrogen, exhibit wave functions representing the internuclear distances, with amplitudes confined to the range $0.1-5$ \AA.
Current state-of-the-art experiments on neutral molecular hydrogen determine vibrational splittings on the order of $10^{-4}$ \wn, or 3 MHz~\cite{Dickenson2013}. Since the lifetimes of ro-vibrational quantum states in H$_2$ are of the order of $10^6$ s~\cite{Black1976}, measurements of vibrational splittings on the order of $10^{14}$ Hz could in principle be possible at more than 20-digit precision, which leaves room for improvement ``at the bottom" of over 10 orders of magnitude, if experimental techniques can be developed accordingly. Similar improvements in theory would make these molecular systems an ideal test ground for constraining or detecting higher dimensions, as well as fifth forces~\cite{Salumbides2013}.
After having performed a 15-digit accuracy calculation on Born-Oppenheimer energies~\cite{Pachucki2010}, calculations of strongly improved accuracy have just been published~\cite{Pachucki2014}, while improved calculations of non-adiabatic corrections are underway~\cite{Pachucki2015}.
Immediate improvements, based on existing technologies, on the experimental accuracies of the dissociation limits in the neutral hydrogen and its isotopomers~\cite{Sprecher2011} and the spectroscopy of HD$^+$~\cite{Karr2014,Tran2013,Schiller2014}, were discussed recently.

\ack
This work was supported by the Netherlands Foundation for Fundamental Research of Matter (FOM) through the program ``Broken Mirrors \& Drifting Constants".
B.~Gato-Rivera and A.N.~Schellekens have been partially supported by funding from the Spanish Ministerio de Economia y Competitividad, Research Project FIS2012-38816, and by the Project CONSOLIDER-INGENIO 2010, Programme CPAN (CSD2007-00042).

\end{document}